 \documentclass[preprint,superscriptaddress,showpacs,prl]{revtex4-1}
\usepackage[dvips]{graphicx}
\usepackage{epsfig}
\begin{document}

\title[Non-local gyrokinetic model of linear ion-temperature-gradient modes]{Non-local gyrokinetic model of linear ion-temperature-gradient modes}
\author{S. Moradi$^{1}$, J. Anderson$^{1}$ and B. Weyssow$^{2}$\\{\it \small $^1$
Department of Applied Physics, Nuclear Engineering, Chalmers
  University of Technology and Euratom-VR Association, G\"oteborg,
  Sweden\\
$^{2}$EFDA-CSU, D-85748 Garching, M\"{u}nchen, Germany}}

\begin{abstract}
A theory of non-local linear ion-temperature-gradient (ITG) drift modes while retaining non-adiabatic electrons is presented, extending the previous work [S. Moradi, et al {\em Phys. Plasmas} {\bf 18}, 062106 (2011)]. A dispersion relation is derived to quantify the effects of the fractional velocity operator in the Fokker-Planck equation modified by temperature gradients and non-adiabatic electrons on the real frequency and growth rate. Solving the dispersion relation,  it is shown here that as the plasma becomes more turbulent, it deviates from a Maxwellian distribution and becomes L\'{e}vy distributed. The resulting L\'{e}vy distribution of the plasma may thus significantly alter the transport. The relative effect of the fractional derivative is larger on the real frequency than on the growth rate of the ITG mode.
\end{abstract}

\maketitle

\section{Introduction}
The high level of anomalous transport in magnetically confined fusion plasmas is still an unresolved issue in the quest for controlled fusion. Furthermore, a deterministic description of intermittent events in plasma turbulence is improper due to the stochastic nature of the transport exhibiting non-local interactions as well as non-Gaussian probability density functions (PDFs). The PDFs of heat and particle flux display uni-modal non-Gaussian features which is the signature of intermittent turbulence with patchy spatial structure that is bursty in time. The turbulent behavior in magnetically confined plasmas is the main ingredient in the anomalously high transport of heat, particles and momentum visible in present days large experiments. One crucial component of the turbulent transport is the so-called ion-temperature-gradient (ITG) driven turbulence. The ITG turbulence is found to be bursty in nature where a significant part of the transport is carried by large avalanche-like events. More specifically, exponential scalings are often observed in the PDF tails in magnetic confinement experiments, and intermittency at the edge strongly influences the overall global particle and heat transport. In particular it may for instance influence the threshold for the high confinement mode (H-mode) in tokamak experiments \cite{connor}. In view of these experimental results, theories built on average transport coefficients and Gaussian statistics fall short in predicting vital transport processes. There is a considerable amount of experimental evidence \cite{Zweben, callen1997, vanmilligen2002, BalescuBook} and recent numerical gyrokinetic \cite{pradalier, villard, mcmillan, sanchez2008} and fluid simulations \cite{negrete2005} that plasma turbulence in tokamaks is highly non-local. A satisfactorily understanding of the non-local signatures as well as the ever-present non-Gaussian PDFs of transport \cite{carreras1996, anderson1, anderson2} found in experiments and numerical simulations is still lacking.

An attractive candidate for explaining the non-local features of ITG turbulence is by inclusion of a fractional velocity operator in the Fokker-Planck (FP) equation \cite{moradipop2011} yielding a non-local description that have non-Gaussian PDFs of heat and particle flux. The fractional operator introduces an inherently non-local description with strongly non-Maxwellian features of the distribution function resulting in significant modification of the transport process. The non-locality is introduced through the integral description of the fractional derivative \cite{zaslavsky, sanchez2006}. There are a number of other phenomenological studies of the effects of fractional derivative models. Using fractional generalizations of the Liouville equation, kinetic descriptions have been developed previously \cite{zaslavsky2002, tarasov2005}. It has been shown that the chaotic dynamics can be described by using the FP equation with coordinate fractional derivatives as a possible tool for the description of anomalous diffusion \cite{zaslavsky1994} and much work has been devoted on investigation of the Langevin equation with L\'{e}vy white noise, see e.g.  Refs. \cite{West1982, metzler1999}, or related fractional FP equation. Furthermore, fractional derivatives have been introduced into the FP framework in a similar manner \cite{chechkin2000, chechkin2002} as the present work however a study on ITG modes is still lacking.

In this paper we introduce the L\'{e}vy statistics into a Langevin equation which yields a fractional FP description. In order to calculate an equilibrium PDF we use a model based on the motion of a charged L\'{e}vy particle in a constant external magnetic field obeying non-Gaussian, L\'{e}vy statistics. This assumption is the natural generalization of the classical example of the motion of a charged Brownian particle with the usual Gaussian statistics. The fractional derivative is represented with the Fourier transform containing a fractional exponent. Here, we extend the work presented in Ref. \cite{moradipop2011} to include the effects of finite temperature gradients and non-adiabatic electrons leading to a fractional description of the non-local effects in ITG turbulent transport in a gyrokinetic framework. We quantify the non-local effects in terms of a modified dispersion relation for linear ITG modes. We have considered a case with constant external magnetic field and a shear-less slab geometry. The characteristics of the ITG modes are fundamentally changed, i.e., the values of the growth rate and real frequency are significantly altered by the order of the fractional derivative $\alpha$. However, the relative effect of the fractional derivative is larger on the real frequency than on the growth rate of the ITG mode. This is different from the results obtained in Ref. \cite{moradipop2011} where the growth rate was increased strongly as the plasma deviated from $\alpha=2$ limit (Maxwellian). We have found that the basis of this difference is due to the different assumptions on the electrons: adiabatic and non-adiabatic electrons are considered in Ref. \cite{moradipop2011} and in the present work, respectively.  

The paper is organized as follows: first we present the mathematical framework of the fractional FP equation (FFPE) which is used to derive a dispersion relation for the ITG modes while retaining the non-local interactions. In the next section, the deviations from a Maxwellian distribution function are investigated and the dispersion relation is solved. We conclude the paper with a results and discussion section.

\section{Fractional Fokker-Planck Equation}
Following Ref.~\cite{moradipop2011}, the FFPE with fractional velocity derivatives in shear-less slab geometry in the presence of a constant external force can be written as:
\begin{eqnarray}\label{eq:1.1}
\frac{\partial F_{s}}{\partial t}+\mathbf{v}\frac{\partial F_{s}}{\partial \mathbf{r}}+\frac{\mathbf{F}}{m_{s}}\frac{\partial F_{s}}{\partial \mathbf{v}}=\nu\frac{\partial }{\partial \mathbf{v}}(\mathbf{v}F_{s})+D\frac{\partial^{\alpha} F_{s}}{\partial |\mathbf{v}|^{\alpha}},
\end{eqnarray}
where $s(=e,i)$ represents the particle species and $0\le\alpha\le 2$. Here, the term $\frac{\partial^{\alpha} F_{s}}{\partial |\mathbf{v}|^{\alpha}}$ is the fractional Riesz derivative. The diffusion coefficient, $D$, is related to the damping term $\nu$, according to a generalized Einstein relation \cite{barkai} 
\begin{eqnarray}\label{eq:1.2}
D=\frac{2^{\alpha-1}T_{\alpha}\nu}{\Gamma(1+\alpha)m_{s}^{\alpha-1}}.
\end{eqnarray}
Here, $T_{\alpha}$ is a generalized temperature, and force $\mathbf{F}$ represents the Lorentz force (due to a constant magnetic field and a zero-averaged electric field) acting on the particles of species $s$ with mass $m_{s}$ and $\Gamma(1+\alpha)$ is the Euler gamma function. The solution of the equation (\ref{eq:1.1}), i.e. the generalized equilibrium distribution, for a general $\alpha$ can be obtained as \cite{moradipop2011}:
\begin{eqnarray}\label{eq:1.3}
F_{s}(\mathbf{r},\mathbf{v})=\frac{n_{s}(\mathbf{r})}{2\pi^{3/2}\sqrt{2\mathcal{D}}} \int \frac{d\mathbf{k}_{\bot}^{v}d\mathbf{k}_{\parallel}^{v}}{(2\pi)^{3/2}}e^{-i(\mathbf{k}_{\bot}^{v}\mathbf{v}_{\bot}+\mathbf{k}_{\parallel}^{v}\mathbf{v}_{\parallel})}e^{-\frac{\mathcal{D}}{\alpha}(|\mathbf{k}^{v}_{\bot}|^{\alpha}+|\mathbf{k}^{v}_{\parallel}|^{\alpha})},
\end{eqnarray}
where
\begin{eqnarray}\label{eq:1.4}
\mathcal{D}=\frac{V_{T,s}^{\alpha}}{\Gamma(1+\alpha)},
\end{eqnarray}
and we have introduced a generalized thermal velocity as 
\begin{eqnarray}\label{eq:1.5}
V_{T,s}^{\alpha}=\frac{2^{\alpha-1}T_{\alpha}}{m_{s}^{\alpha-1}}.
\end{eqnarray}
Using the generalized equilibrium distribution expressed in equation (\ref{eq:1.3}), we will now quantify the non-local effects on drift waves induced by the fractional differential operator by determining the dispersion relation for ITG driven drift modes. We start by formulating the linearized gyro-kinetic theory where the particle distribution function, averaged over gyro-phase is of the form (see Ref. \cite{Balescu1991})
\begin{eqnarray}\label{eq:2.1}
f_{s}(\mathbf{r},\mathbf{v})=F_{s}(\mathbf{r},\mathbf{v})+(2\pi)^{-4}\times\int\int d\mathbf{k}\;d\omega\exp(i\mathbf{k}\cdot\mathbf{r}-i\omega t)\delta f^{s}_{\mathbf{k},\omega}(\mathbf{v}).
\end{eqnarray} 
We assume that the turbulence is purely electrostatic and neglect magnetic field fluctuations $(\delta \mathbf{B}=0)$. For small deviations from the local equilibrium we find the linearized gyro-kinetic equation of the form
\begin{eqnarray}\label{eq:2.2}
(\partial_t+i k_{\parallel}v_{\parallel})\delta f^{s}_{\mathbf{k}}(v_{\parallel},v_{\bot},t)=i[\frac{c}{B}k_y\nabla_{x}+\frac{e_s}{m_s}k_{\parallel}\partial_{\parallel}]F_{s}(x,v_{\parallel},v_{\bot})J_{0}(|\Omega_{s}|^{-1}k_{\bot}v_{\bot})\delta \phi_{\mathbf{k}}(t).
\end{eqnarray} 
Here $\partial_{\parallel}=\partial/\partial v_{\parallel}$. Evaluating explicitly the derivatives of the distribution function in equation (\ref{eq:1.3}), we obtain the following relations:
\begin{eqnarray}\label{eq:2.3}
\frac{c}{B}k_y\nabla_{x}F_{s}(x,\mathbf{v})=\frac{e_s}{T_{s,\alpha}}\omega_{*\mathbf{k}}^s[\frac{d\;ln\;n_s(x)}{d x}-\frac{1}{2}\frac{d\;ln\;T_{s,\alpha}(x)}{d x}]F_{s}(x,\mathbf{v})+\frac{e_s}{T_{s,\alpha}}\omega_{*\mathbf{k}}^s\times\nonumber\\
\{\frac{n_{s}(x)}{2\pi^{3/2}\sqrt{2\mathcal{D}}} \int \frac{d\mathbf{k}_{\bot}^{v}d\mathbf{k}_{\parallel}^{v}}{(2\pi)^{3/2}}[-\frac{\mathcal{D}}{\alpha}(|\mathbf{k}^{v}_{\bot}|^{\alpha}+|\mathbf{k}^{v}_{\parallel}|^{\alpha})\frac{d\;ln\;T_{s,\alpha}(x)}{d x}]e^{-i(\mathbf{k}_{\bot}^{v}\mathbf{v}_{\bot}+\mathbf{k}_{\parallel}^{v}\mathbf{v}_{\parallel})}e^{-\frac{\mathcal{D}}{\alpha}(|\mathbf{k}^{v}_{\bot}|^{\alpha}+|\mathbf{k}^{v}_{\parallel}|^{\alpha})}\},
\end{eqnarray}
where $\frac{d\;ln\;A(x)}{d x}=\frac{1}{A(x)}\frac{dA(x)}{d x}$, and
\begin{eqnarray}\label{eq:2.4}
\frac{e_s}{m_s}k_{\parallel}\partial_{\parallel}F_{s}(x,\mathbf{v})=\frac{e_s}{m_s}k_{\parallel}\frac{n_{s}(x)}{2\pi^{3/2}\sqrt{2\mathcal{D}}} \int \frac{d\mathbf{k}_{\bot}^{v}d\mathbf{k}_{\parallel}^{v}}{(2\pi)^{3/2}}(-i\mathbf{k}_{\parallel}^{v})e^{-i(\mathbf{k}_{\bot}^{v}\mathbf{v}_{\bot}+\mathbf{k}_{\parallel}^{v}\mathbf{v}_{\parallel})}e^{-\frac{\mathcal{D}}{\alpha}(|\mathbf{k}^{v}_{\bot}|^{\alpha}+|\mathbf{k}^{v}_{\parallel}|^{\alpha})},
\end{eqnarray}
where $\omega_{*\mathbf{k}}^s=\frac{cT_{s}}{e_{s}B}k_{y}$, and we assumed that the space dependence of $F_{s}$ is only in the $x$ direction perpendicular to the magnetic field as well as for the density gradient. In the equation above, $J_{0}$ is the Bessel function of order zero, $v_{\parallel}$ is the parallel velocity, $v_{\bot}\equiv (v_{x}^{2}+v_{y}^{2})^{1/2}$ is the perpendicular velocity and hence we write the total speed as $v=(v_{\bot}^{2}+v_{\parallel}^{2})^{1/2}$. The linearized gyro-kinetic equation could be further Laplace transformed. The Fourier-Laplace transform of the fluctuating electrostatic potential is
\begin{eqnarray}\label{eq:2.5}
\delta\phi_{\mathbf{k},\omega}=\int_{0}^{\infty} dt e^{i\omega t}\delta \phi_{\mathbf{k}}(t).
\end{eqnarray}
Similar formula defines the Fourier-Laplace transform of $\delta f_{\mathbf{k},\omega}$. Therefore the Fourier-Laplace transformed gyro-kinetic equation (\ref{eq:2.2}) is
\begin{eqnarray}\label{eq:2.6}
-i(\omega- k_{\parallel}v_{\parallel})\delta f^{s}_{\mathbf{k}}(v_{\parallel},v_{\bot},t)=-\Delta^{s}_{\mathbf{k},\omega}(v_{\parallel},v_{\bot})\delta\phi_{\mathbf{k},\omega}+\delta f^{s}_{\mathbf{k}}(v_{\parallel},v_{\bot},0).
\end{eqnarray} 
Its solution is
\begin{eqnarray}\label{eq:2.7}
\delta f^{s}_{\mathbf{k},\omega}(v_{\parallel},v_{\bot})=\mathcal{G}^{s}_{\mathbf{k},\omega}(v_{\parallel},v_{\bot})\{-\Delta^{s}_{\mathbf{k},\omega}(v_{\parallel},v_{\bot})\delta\phi_{\mathbf{k},\omega}+\delta f^{s}_{\mathbf{k}}(v_{\parallel},v_{\bot},0)\},
\end{eqnarray}
where the operator
\begin{eqnarray}\label{eq:2.8}
\mathcal{G}^{s}_{\mathbf{k},\omega}(v_{\parallel},v_{\bot})=\frac{1}{-i(\omega-k_{\parallel}v_{\parallel})}
\end{eqnarray}
is the unperturbed propagator of the gyro-kinetic equation, and we have introduced the function $\Delta^{s}_{\mathbf{k},\omega}(v_{\parallel},v_{\bot})$ as
\begin{eqnarray}\label{eq:2.9}
&&\Delta^{s}_{\mathbf{k},\omega}(v_{\parallel},v_{\bot})=\nonumber\\
&&-i\frac{e_s}{T_{s,\alpha}}\omega_{*\mathbf{k}}^s[\frac{d\;ln\; n_s(x)}{d x}-\frac{1}{2}\frac{d\;ln\;T_{s,\alpha}(x)}{d x}]F_{s}(x,\mathbf{v})J_{0}(|\Omega_{s}|^{-1}k_{\bot}v_{\bot})+i\frac{e_s}{T_{s,\alpha}}\omega_{*\mathbf{k}}^s\times\nonumber\\
&&\left\{\frac{n_{s}(x)}{2\pi^{3/2}\sqrt{2\mathcal{D}}} \int \frac{d\mathbf{k}_{\bot}^{v}d\mathbf{k}_{\parallel}^{v}}{(2\pi)^{3/2}}\left[\frac{\mathcal{D}}{\alpha}(|\mathbf{k}^{v}_{\bot}|^{\alpha}+|\mathbf{k}^{v}_{\parallel}|^{\alpha})\frac{d\;ln\;T_{s,\alpha}(x)}{d x}\right]e^{-i(\mathbf{k}_{\bot}^{v}\mathbf{v}_{\bot}+\mathbf{k}_{\parallel}^{v}\mathbf{v}_{\parallel})}e^{-\frac{\mathcal{D}}{\alpha}(|\mathbf{k}^{v}_{\bot}|^{\alpha}+|\mathbf{k}^{v}_{\parallel}|^{\alpha})}\right\}J_{0}(|\Omega_{s}|^{-1}k_{\bot}v_{\bot})\nonumber\\
&&+i\frac{e_s}{T_{s,\alpha}}\left[\frac{T_{s,\alpha}}{m_s}k_{\parallel}\frac{n_{s}(x)}{2\pi^{3/2}\sqrt{2\mathcal{D}}} \int \frac{d\mathbf{k}_{\bot}^{v}d\mathbf{k}_{\parallel}^{v}}{(2\pi)^{3/2}}(i\mathbf{k}_{\parallel}^{v})e^{-i(\mathbf{k}_{\bot}^{v}\mathbf{v}_{\bot}+\mathbf{k}_{\parallel}^{v}\mathbf{v}_{\parallel})}e^{-\frac{\mathcal{D}}{\alpha}(|\mathbf{k}^{v}_{\bot}|^{\alpha}+|\mathbf{k}^{v}_{\parallel}|^{\alpha})}\right]J_{0}(|\Omega_{s}|^{-1}k_{\bot}v_{\bot}).
\end{eqnarray}
Here, the wave vector perpendicular to magnetic field is $k_{\bot}=(k^2_{x}+k^2_{y})^{1/2}$. The gyro-kinetic Equation (\ref{eq:2.1}) is complemented with Poisson equation for the electric potential. For fluctuations with wave vectors much smaller than the Debye wave vector, the Poisson equation becomes the quasi-neutrality condition
\begin{eqnarray}\label{eq:2.10}
\sum_{s} e_{s}\delta n^{s}_{\mathbf{k},\omega}=0,
\end{eqnarray}
where the density fluctuation is related to the distribution function through
\begin{eqnarray}\label{eq:2.11}
\delta n^{s}_{\mathbf{k},\omega}=-\frac{e_{s}}{T_{s}}n_{s}\delta\phi_{\mathbf{k},\omega} + \int d\mathbf{v}
J_{0}(|\Omega_{s}|^{-1}k_{\bot}v_{\bot})\delta f^{s}_{\mathbf{k},\omega}(v_{\parallel},v_{\bot}).
\end{eqnarray}
In the above equation we have separated the adiabatic response (first term on the right hand side) from the non-adiabatic response (second term on the right hand side). We have to keep in mind that the density $n_{s}$ coming from the $F_{s}(x,\mathbf{v})$ in the adiabatic response is also given by Equation (\ref{eq:1.3}) and for a general $0\le\alpha\le2$ the adiabatic response can be different than that calculated by Maxwellian distribution. Using the quasi-neutrality condition (\ref{eq:2.4}) we find the dispersion equation which determines the eigenfrequencies as a function of the wave vector, $\omega=\omega(\mathbf{k})=\omega_{r}(\mathbf{k})+i\gamma(\mathbf{k})$. In the simplest case we consider a plasma consisting of electrons and a single species of singly charged ions with equal temperatures. For the density fluctuation therefore we have
\begin{eqnarray}\label{eq:2.12}
\delta n^{s}_{\mathbf{k},\omega}=-n_{s}(x)\frac{e_{s}}{T_{s}}\delta\phi_{\mathbf{k},\omega}[M^{ad,s}+M^{s}_{\mathbf{k},\omega}].
\end{eqnarray}
Therefore, the dispersion equation as in the Ref. \cite{Balescu1991} is
\begin{eqnarray}\label{eq:2.13}
M^{ad,e}+M^{e}_{\mathbf{k},\omega}=-M^{ad,i}-M^{i}_{\mathbf{k},\omega},
\end{eqnarray}
where
\begin{eqnarray}\label{eq:2.14}
M^{ad,s}=\int d\mathbf{v}
\frac{1}{2\pi^{3/2}(\Gamma(1+\alpha))^{-1/2}\sqrt{2V_{T,s}^{\alpha}}} 
\int \frac{d\mathbf{k}_{\bot}^{v}d\mathbf{k}_{\parallel}^{v}}{(2\pi)^{3/2}}e^{-i(\mathbf{k}_{\bot}^{v}\mathbf{v}_{\bot}+\mathbf{k}_{\parallel}^{v}\mathbf{v}_{\parallel})}e^{-\frac{V_{T,s}^{\alpha}}{\Gamma(1+\alpha)\alpha}(|\mathbf{k}^{v}_{\bot}|^{\alpha}+|\mathbf{k}^{v}_{\parallel}|^{\alpha})},
\end{eqnarray}
gives the adiabatic contribution, and
\begin{eqnarray}\label{eq:2.15}
M^{s}_{\mathbf{k},\omega}=\frac{1}{n_{s}(x)}\int d\mathbf{v}\mathcal{G}^{s}_{\mathbf{k},\omega}(v_{\parallel},v_{\bot})\Delta^{s}_{\mathbf{k},\omega}(v_{\parallel},v_{\bot})J_{0}(|\Omega_{s}|^{-1}k_{\bot}v_{\bot})=\nonumber\\
-\omega_{*\mathbf{k}}^s[\frac{d\;ln\; n_s(x)}{d x}-\frac{1}{2}\frac{d\;ln\;T_{s,\alpha}(x)}{d x}]\int d\mathbf{v}\frac{J_{0}^2(|\Omega_{s}|^{-1}k_{\bot}v_{\bot})}{\omega-k_{\parallel}v_{\parallel}}\{\frac{1}{2\pi^{3/2}(\Gamma(1+\alpha))^{-1/2}\sqrt{2V_{T,s}^{\alpha}}} \times\nonumber\\
\int \frac{d\mathbf{k}_{\bot}^{v}d\mathbf{k}_{\parallel}^{v}}{(2\pi)^{3/2}}e^{-i(\mathbf{k}_{\bot}^{v}\mathbf{v}_{\bot}+\mathbf{k}_{\parallel}^{v}\mathbf{v}_{\parallel})}e^{-\frac{V_{T,s}^{\alpha}}{\Gamma(1+\alpha)\alpha}(|\mathbf{k}^{v}_{\bot}|^{\alpha}+|\mathbf{k}^{v}_{\parallel}|^{\alpha})}\}+\nonumber\\
\omega_{*\mathbf{k}}^s\int d\mathbf{v}\frac{J_{0}^2(|\Omega_{s}|^{-1}k_{\bot}v_{\bot})}{\omega-k_{\parallel}v_{\parallel}}\{\frac{1}{2\pi^{3/2}(\Gamma(1+\alpha))^{-1/2}\sqrt{2V_{T,s}^{\alpha}}} \times\nonumber\\
\int \frac{d\mathbf{k}_{\bot}^{v}d\mathbf{k}_{\parallel}^{v}}{(2\pi)^{3/2}}[\frac{V_{T,s}^{\alpha}}{\Gamma(1+\alpha)\alpha}(|\mathbf{k}^{v}_{\bot}|^{\alpha}+|\mathbf{k}^{v}_{\parallel}|^{\alpha})\frac{d\;ln\;T_{s,\alpha}(x)}{d x}]e^{-i(\mathbf{k}_{\bot}^{v}\mathbf{v}_{\bot}+\mathbf{k}_{\parallel}^{v}\mathbf{v}_{\parallel})}e^{-\frac{V_{T,s}^{\alpha}}{\Gamma(1+\alpha)\alpha}(|\mathbf{k}^{v}_{\bot}|^{\alpha}+|\mathbf{k}^{v}_{\parallel}|^{\alpha})}\}+\nonumber\\
\frac{T_{s,\alpha}}{m_s}k_{\parallel}\int d\mathbf{v}\frac{J_{0}^2(|\Omega_{s}|^{-1}k_{\bot}v_{\bot})}{\omega-k_{\parallel}v_{\parallel}}\{\frac{1}{2\pi^{3/2}(\Gamma(1+\alpha))^{-1/2}\sqrt{2V_{T,s}^{\alpha}}} \times\nonumber\\
\int \frac{d\mathbf{k}_{\bot}^{v}d\mathbf{k}_{\parallel}^{v}}{(2\pi)^{3/2}}[i\mathbf{k}_{\parallel}^{v}]e^{-i(\mathbf{k}_{\bot}^{v}\mathbf{v}_{\bot}+\mathbf{k}_{\parallel}^{v}\mathbf{v}_{\parallel})}e^{-\frac{V_{T,s}^{\alpha}}{\Gamma(1+\alpha)\alpha}(|\mathbf{k}^{v}_{\bot}|^{\alpha}+|\mathbf{k}^{v}_{\parallel}|^{\alpha})}\}\nonumber\\
\end{eqnarray}
gives the non-adiabatic contribution. 

The analytical solutions for integrals over $\mathbf{k}^{v}$ with an arbitrary $\alpha$ in the Equations (\ref{eq:2.14}) and (\ref{eq:2.15}) requires rather tedious calculations. Instead we consider an infinitesimal deviation of the form $\alpha=2-\epsilon$, where $0\le\epsilon\ll 2$ and expand the terms depending on $\alpha$ in the Equation (\ref{eq:2.14}) around $\epsilon=0$ as follows
\begin{eqnarray}\label{eq:2.16}
\frac{1}{2\pi^{3/2}(\Gamma(1+\alpha))^{-1/2}\sqrt{2V_{T,s}^{\alpha}}} e^{-\frac{V_{T,s}^{\alpha}}{\Gamma(1+\alpha)\alpha}(|\mathbf{k}^{v}_{\bot}|^{\alpha}+|\mathbf{k}^{v}_{\parallel}|^{\alpha})}=\nonumber\\
\frac{e^{-\frac{V^2_{T,s}}{4}(|\mathbf{k}^{v}_{\bot}|^{2}+|\mathbf{k}^{v}_{\parallel}|^{2})}}{2\pi^{3/2}V_{T,s}}+\Lambda(\mathbf{k}_{\bot}^{v},\mathbf{k}_{\parallel}^{v})\epsilon+\mathcal{O}[\epsilon^2],
\end{eqnarray}
where
\begin{eqnarray}\label{eq:2.17}
\Lambda(\mathbf{k}_{\bot}^{v},\mathbf{k}_{\parallel}^{v})=\frac{e^{-\frac{V^2_{T,s}}{4}(|\mathbf{k}^{v}_{\bot}|^{2}+|\mathbf{k}^{v}_{\parallel}|^{2})}}{8\pi^{3/2}V_{T,s}}\{-3+2{\gamma_E}+2{\log}[V_{T,s}]-2V_{T,s}^2[|\mathbf{k}^{v}_{\bot}|^{2}+|\mathbf{k}^{v}_{\parallel}|^{2}]+\nonumber\\
{\gamma_E}V_{T,s}^2(|\mathbf{k}^{v}_{\bot}|^{2}+|\mathbf{k}^{v}_{\parallel}|^{2})+V_{T,s}^2(|\mathbf{k}^{v}_{\bot}|^{2}{\log}[|\mathbf{k}^{v}_{\bot}|^{2}]+|\mathbf{k}^{v}_{\parallel}|^{2}{\log}[|\mathbf{k}^{v}_{\parallel}|^{2}])+V_{T,s}^2{\log}[V_{T,s}](|\mathbf{k}^{v}_{\bot}|^{2}+|\mathbf{k}^{v}_{\parallel}|^{2})\}\nonumber\\
\end{eqnarray}
and in Equation (\ref{eq:2.15}) the expansion for the second term on the RHS gives
\begin{eqnarray}\label{eq:2.18}
\frac{\frac{V_{T,s}^{\alpha}}{\Gamma(1+\alpha)\alpha}(|\mathbf{k}^{v}_{\bot}|^{\alpha}+|\mathbf{k}^{v}_{\parallel}|^{\alpha})}{2\pi^{3/2}(\Gamma(1+\alpha))^{-1/2}\sqrt{2V_{T,s}^{\alpha}}} e^{-\frac{V_{T,s}^{\alpha}}{\Gamma(1+\alpha)\alpha}(|\mathbf{k}^{v}_{\bot}|^{\alpha}+|\mathbf{k}^{v}_{\parallel}|^{\alpha})}=\nonumber\\
\frac{e^{-\frac{V^2_{T,s}}{4}(|\mathbf{k}^{v}_{\bot}|^{2}+|\mathbf{k}^{v}_{\parallel}|^{2})}V_{T,s}(|\mathbf{k}^{v}_{\bot}|^{2}+|\mathbf{k}^{v}_{\parallel}|^{2})}{8\pi^{3/2}}+\Sigma(\mathbf{k}_{\bot}^{v},\mathbf{k}_{\parallel}^{v})\epsilon+\mathcal{O}[\epsilon^2],
\end{eqnarray}
where
\begin{eqnarray}\label{eq:2.19}
\Sigma(\mathbf{k}_{\bot}^{v},\mathbf{k}_{\parallel}^{v})=\frac{e^{-\frac{V^2_{T,s}}{4}(|\mathbf{k}^{v}_{\bot}|^{2}+|\mathbf{k}^{v}_{\parallel}|^{2})}V_{T,s}}{32\pi^{3/2}}\{5(|\mathbf{k}^{v}_{\bot}|^{2}+|\mathbf{k}^{v}_{\parallel}|^{2})-(2{\gamma_E}+2{\log}[V_{T,s}])(|\mathbf{k}^{v}_{\bot}|^{2}+|\mathbf{k}^{v}_{\parallel}|^{2})\nonumber\\
-4(|\mathbf{k}^{v}_{\bot}|^{2}{\log}[|\mathbf{k}^{v}_{\bot}|]+|\mathbf{k}^{v}_{\parallel}|^{2}{\log}[|\mathbf{k}^{v}_{\parallel}|])+(-2V_{T,s}+\gamma_{E}V_{T,s}^2+{\log}[V_{T,s}]V_{T,s}^2)(|\mathbf{k}^{v}_{\bot}|^{4}+|\mathbf{k}^{v}_{\parallel}|^{4})\nonumber\\
(-4V_{T,s}+2\gamma_{E}V_{T,s}^2+2{\log}[V_{T,s}]V_{T,s}^2)(|\mathbf{k}^{v}_{\bot}|^{2}|\mathbf{k}^{v}_{\parallel}|^{2})+V_{T,s}^2(|\mathbf{k}^{v}_{\bot}|^{4}{\log}[|\mathbf{k}^{v}_{\bot}|]+|\mathbf{k}^{v}_{\parallel}|^{4}{\log}[|\mathbf{k}^{v}_{\parallel}|])+\nonumber\\
V_{T,s}^2(|\mathbf{k}^{v}_{\bot}|^{2}|\mathbf{k}^{v}_{\parallel}|^{2})({\log}[|\mathbf{k}^{v}_{\bot}|]+{\log}[|\mathbf{k}^{v}_{\parallel}|])\}.
\end{eqnarray}
Here, we have used the Euler-Mascheroni constant $\gamma_E \approx 0.57721$. 

Inserting the zeroth order terms in $\epsilon$ from the expansion (\ref{eq:2.16}) into Equation (\ref{eq:2.14}) will produce the Maxwellian adiabatic response
\begin{eqnarray}\label{eq:2.20}
M^{ad,s}=1,
\end{eqnarray}
and by inserting the  zeroth order terms in $\epsilon$ from the expansion (\ref{eq:2.17}) into Equation (\ref{eq:2.15}) will produce the Maxwellian non-adiabatic response
\begin{eqnarray}\label{eq:2.21}
M^{s}_{\mathbf{k},\omega}=\frac{2}{\sqrt{\pi}V_{T,s}^{3}}\int_{-\infty}^{\infty} dv_{\parallel}\int_{0}^{\infty} dv_{\bot}v_{\bot}\frac{k_{\parallel}v_{\parallel}-\omega^{s,T}_{*k}(v_{\parallel},v_{\bot})}{-\omega+k_{\parallel}v_{\parallel}}J_{0}^2(|\Omega_{s}|^{-1}k_{\bot}v_{\bot})e^{-\frac{v_{\parallel}^2+v_{\bot}^2}{V_{T,s}^2}},
\end{eqnarray}
where
\begin{eqnarray}\label{eq:2.22}
\omega^{s,T}_{*k}(v_{\parallel},v_{\bot})=\omega_{*\mathbf{k}}^s[\frac{d\;ln\;n_{s}(x)}{dx}+(\frac{v_{\parallel}^2+v_{\bot}^2}{V_{T,s}^2}-\frac{3}{2})\frac{d\;ln\;T_{s}(x)}{dx}].
\end{eqnarray}

By using the expansion defined by the expressions (\ref{eq:2.16}) and (\ref{eq:2.18}) to first order in $\epsilon$ from Equations (\ref{eq:2.14}) and (\ref{eq:2.15}), the adiabatic and non-adiabatic parts of the dispersion relation $M^{ad,s}$ and $M^{s}_{\mathbf{k},\omega}$ are as follows
\begin{eqnarray}\label{eq:3.8}
M^{ad,s}=1+(2\pi\int_{-\infty}^{\infty}dv_{\parallel}\int_{0}^{\infty}dv_{\bot}v_{\bot}\times\nonumber\\
\int \frac{d\mathbf{k}_{\bot}^{v}d\mathbf{k}_{\parallel}^{v}}{(2\pi)^{3/2}}e^{-i(\mathbf{k}_{\bot}^{v}\mathbf{v}_{\bot}+\mathbf{k}_{\parallel}^{v}\mathbf{v}_{\parallel})}\Lambda(\mathbf{k}_{\parallel}^{v},\mathbf{k}_{\bot}^{v}))\epsilon=1+\epsilon W^{ad,s}.\nonumber\\
\end{eqnarray}
and
\begin{eqnarray}\label{eq:3.9}
M^{s}_{\mathbf{k},\omega}=\frac{2}{\sqrt{\pi}V_{T,s}^{3}}\int_{-\infty}^{\infty} dv_{\parallel}\int_{0}^{\infty} dv_{\bot}v_{\bot}\frac{k_{\parallel}v_{\parallel}-\omega^{s,T}_{*k}(v_{\parallel},v_{\bot})}{-\omega+k_{\parallel}v_{\parallel}}J_{0}^2(|\Omega_{s}|^{-1}k_{\bot}v_{\bot})e^{-\frac{v_{\parallel}^2+v_{\bot}^2}{V_{T,s}^2}}+\epsilon\nonumber\\
\{-2\pi\omega_{*\mathbf{k}}^s[\frac{d\;ln\; n_s(x)}{d x}-\frac{1}{2}\frac{d\;ln\;T_{s}(x)}{d x}]\int_{-\infty}^{\infty}dv_{\parallel}\int_{0}^{\infty}dv_{\bot}v_{\bot}\frac{J_{0}^2(|\Omega_{s}|^{-1}k_{\bot}v_{\bot})}{\omega-k_{\parallel}v_{\parallel}}\times\nonumber\\
\int \frac{d\mathbf{k}_{\bot}^{v}d\mathbf{k}_{\parallel}^{v}}{(2\pi)^{3/2}}e^{-i(\mathbf{k}_{\bot}^{v}\mathbf{v}_{\bot}+\mathbf{k}_{\parallel}^{v}\mathbf{v}_{\parallel})}\Lambda(\mathbf{k}_{\bot}^{v},\mathbf{k}_{\parallel}^{v})+\nonumber\\
2\pi\omega_{*\mathbf{k}}^s \frac{d\;ln\;T_{s}(x)}{d x}\int_{-\infty}^{\infty}dv_{\parallel}\int_{0}^{\infty}dv_{\bot}v_{\bot}\frac{J_{0}^2(|\Omega_{s}|^{-1}k_{\bot}v_{\bot})}{\omega-k_{\parallel}v_{\parallel}}\times\nonumber\\
\int \frac{d\mathbf{k}_{\bot}^{v}d\mathbf{k}_{\parallel}^{v}}{(2\pi)^{3/2}}e^{-i(\mathbf{k}_{\bot}^{v}\mathbf{v}_{\bot}+\mathbf{k}_{\parallel}^{v}\mathbf{v}_{\parallel})}\Sigma(\mathbf{k}_{\bot}^{v},\mathbf{k}_{\parallel}^{v})+\nonumber\\
2\pi\frac{T_{s}}{m_s}k_{\parallel}\int_{-\infty}^{\infty}dv_{\parallel}\int_{0}^{\infty}dv_{\bot}v_{\bot}\frac{J_{0}^2(|\Omega_{s}|^{-1}k_{\bot}v_{\bot})}{\omega-k_{\parallel}v_{\parallel}}\times\nonumber\\
\int \frac{d\mathbf{k}_{\bot}^{v}d\mathbf{k}_{\parallel}^{v}}{(2\pi)^{3/2}}e^{-i(\mathbf{k}_{\bot}^{v}\mathbf{v}_{\bot}+\mathbf{k}_{\parallel}^{v}\mathbf{v}_{\parallel})}(i\mathbf{k}_{\parallel})\Lambda(\mathbf{k}_{\bot}^{v},\mathbf{k}_{\parallel}^{v})\}=N^{s}_{\mathbf{k},\omega}+\epsilon W^{s}_{\mathbf{k},\omega}.
\end{eqnarray}


\section{Dispersion Equation}
We will now turn our attention to the problem of solving the dispersion relation described by Equation (\ref{eq:2.13}). In order to solve this dispersion equation we use  the method described in Ref.~\cite{moradipop2011}, where the dispersion relation is in the form
\begin{eqnarray}\label{eq:3.1}
(1+N^{e}_{\mathbf{k},\omega})+\epsilon_e (W^{ad,e}+W^{e}_{\mathbf{k},\omega})=-(1+N^{i}_{\mathbf{k},\omega})-\epsilon_i (W^{ad,i}+W^{i}_{\mathbf{k},\omega}).
\end{eqnarray}
Note that we have expanded in $\epsilon_e$ and $\epsilon_i$ for electrons and ions, respectively and that there exist a relation between the two see Ref. \cite{moradipop2011}. The first terms on the right and left hand sides generate the usual contributions to the dispersion equation as in Ref. \cite{Balescu1991} and the terms proportional to $\epsilon$ generate the non-Maxwellian contributions. For the non-adiabatic Maxwellian response we have
\begin{eqnarray}\label{eq:3.2}
N^{s}_{\mathbf{k},\omega}=\frac{2}{\sqrt{\pi}}\int_{-\infty}^{\infty} dw\int_{0}^{\infty} du u[\frac{w-\bar{\omega}^{s,T}_{*k}(u,w)}{w-\bar{\omega}}]J_{0}^2(b_{s}u)e^{-(u^2+w^2)},
\end{eqnarray}
with
\begin{eqnarray}\label{eq:3.3}
\bar{\omega}_{*k}^{sT}(u,w)=\bar{\omega}_{*\mathbf{k}}^s[1+(u^2+w^2-\frac{3}{2})\eta_s].
\end{eqnarray}
Here, $b_{s}=k_{\bot}V_{T,s}/\Omega_{s}$, $\{w,u\}=\{v_{\parallel}/V_{T,s},v_{\bot}/V_{T,s}\}$, we have introduced the following notation $L_A=\frac{d\;ln\;A(x)}{d x}$, $\eta_s=L_T/L_n$ and $\omega_{*\mathbf{k}}^s=\frac{cT_{s}}{e_{s}B}k_{y}/L_n$. Bar denotes normalization to $\mathbf{k}_{\parallel}V_{T,s}$.
The effects of the fractional velocity derivative can result in the non-Maxwellian contribution of the form
\begin{eqnarray}\label{eq:3.4}
W^{s}_{\mathbf{k},\omega}=\frac{2}{\sqrt{\pi}}\int_{-\infty}^{\infty} dw\int_{0}^{\infty} du u[\frac{w\Upsilon(u,w)-\bar{\Omega}_{*k}^{sT}(u,w)}{w-\bar{\omega}}]J_{0}^2(b_{s}u)e^{-(u^2+w^2)},
\end{eqnarray}
where
\begin{eqnarray}\label{eq:3.5}
\bar{\Omega}_{*k}^{sT}(u,w)=\bar{\omega}_{*\mathbf{k}}^s[1-\frac{1}{2}\eta_s]\Phi(u,w)-\bar{\omega}_{*\mathbf{k}}^s \eta_s\Psi(u,w).
\end{eqnarray}
The functions $\Phi(u,w)$, $\Psi(u,w)$ and $\Upsilon(u,w)$ are given in Appendix A.

\section{Results and discussion}
In this section we present the solutions of the dispersion Equation (\ref{eq:3.1}) using Equations (\ref{eq:3.2}) and (\ref{eq:3.4}). We can find an expression for $\epsilon_{e}$ as:
\begin{eqnarray}\label{eq:4.1}
\epsilon_{e}=-\frac{2+N^{e}_{\mathbf{k},\omega}+N^{i}_{\mathbf{k},\omega}}{67.32+W^{e}_{\mathbf{k},\omega}+1.42W^{i}_{\mathbf{k},\omega}}.
\end{eqnarray}
Here, we have used the results shown in Ref.~\cite{moradipop2011}: $\epsilon_{i}=1.42\epsilon_{e}$, $W^{ad}_{e}=33.724$, and $W^{ad}_{i}=23.6591$, $V_{T,e}=5.93\;10^{9} [cm/s]$, $V_{T,i}=1.38\;10^{8} [cm/s]$, and $b_{i}=0.42$. We normalize all the frequencies to $|k_{\parallel}|V_{Te}$ and we solve Equation (\ref{eq:4.1}) for given values of $\gamma$ and $\omega$ where $\bar{\omega}=\omega_r+i \gamma$ corresponding to the real and imaginary (also called growth rate) parts of the eigenvalue. 

\begin{figure}[tbp]
\begin{center}
\epsfig{figure=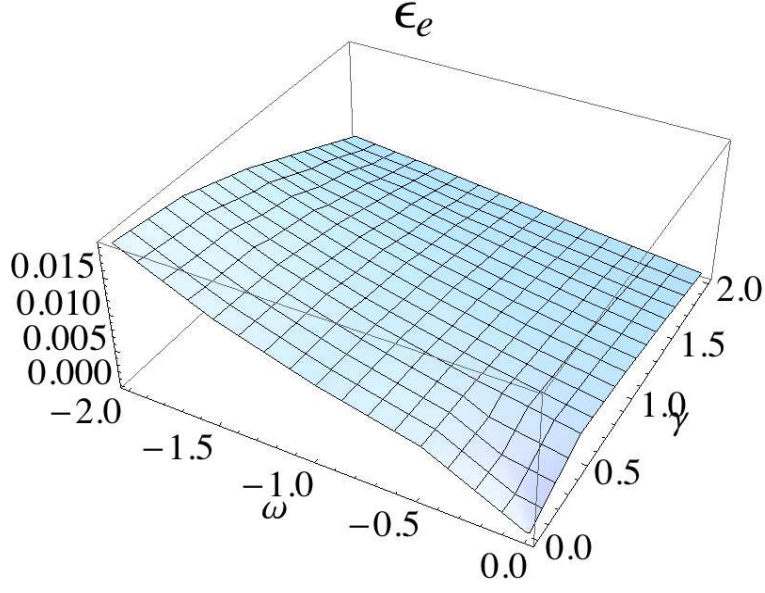, width=10cm,height=8cm,clip=}
\end{center}
\caption{$\epsilon_e$ as function of $\omega$ and $\gamma$, where the frequencies are normalized to $|k_{\parallel}|V_{Te}$. }
\label{fig1}
\end{figure}

Figure \ref{fig1} shows the deviation factor $\epsilon_e$, as defined in equation (\ref{eq:4.1}) calculated for given values of $\gamma$ and $\omega$ with $\eta_i=5$ and $\eta_e=0$. As seen in this figure, the deviation factor increases as the frequency and growth rate of the ITG mode increase. These results are in agreement with results given in Ref.~\cite{moradipop2011}, where it was shown that as the growth rate increases, e.g. the plasma become more turbulent, the plasma starts to deviate from a Maxwellian, and becomes L\'{e}vy distributed. This qualitative behavior is observed for all relevant values of the temperature gradient through the parameter $\eta_i$. Note that the relative effect on the real frequency is larger compared to the effect on the growth rate. This behavior is different from the results shown in Ref. \cite{moradipop2011} where the main effect of the deviation of the plasma from Maxwellian was observed on the growth rate of the density gradient mode. From our findings we expect that the basis of this difference is due to the difference in the assumptions on the electron dynamics: adiabatic or non-adiabatic electrons were assumed in Ref. \cite{moradipop2011} and here, respectively. However, these results in agreement with \cite{moradipop2011} suggest that as the plasma becomes more turbulent, it starts to deviate from a Maxwellian distribution and becomes L\'{e}vy distributed. The resulting L\'{e}vy distribution of the plasma may thus significantly alter the transport. Therefore, the impact of the redistribution of the plasma with different statistical properties has to be taken into account when calculating the transport effects.



\section{Acknowledgements}
This work was funded by the European Communities under Association Contract between EURATOM and
{\em Vetenskapsr{\aa}det}. 

\appendix
\section*{Appendix A}
\setcounter{section}{1}
The functions $\Phi(u,w)$, $\Psi(u,w)$ and $\Upsilon(u,w)$ are defined as followes:
\begin{eqnarray}\label{eq:3.6}
&&\Phi(u,w)=-\frac{i}{8|u|}\{u\mathbf{Erfi}[|u|]\nonumber\\
&&((-1+3w^2)(-2\gamma_{E}+2{\log}[V_{T,s}])+e^{w^2}{_1 F_1}^{(1,0,0)}[\frac{3}{2},\frac{1}{2},-w^2])\nonumber\\
&&-i|u|(14-8\gamma_{E}-8u^2+4\gamma_{E}u^2-8w^2+4\gamma_{E}w^2-4{\log}[V_{T,s}]\nonumber\\
&&-2e^{u^2}{_1 F_1}^{(1,0,0)}[\frac{3}{2},\frac{1}{2},-u^2]-2e^{w^2}{_1 F_1}^{(1,0,0)}[\frac{3}{2},\frac{1}{2},-w^2]\nonumber\\
&&-i\mathbf{Erfi}[|u|]((-1+2w^2)(-2+\gamma_{E}+2{\log}[V_{T,s}])+e^{w^2}{_1 F_1}^{(1,0,0)}[\frac{3}{2},\frac{1}{2},-w^2]))\},
\end{eqnarray}
and
\begin{eqnarray}\label{eq:3.7}
&&\Psi(u,w)=\frac{1}{48|u|}\{i u \mathbf{Erfi}[|u|]\nonumber\\
&&((3-12w^2+4w^4)(-8+3\gamma_{E}+6{\log}[V_{T,s}])-9e^{w^2}{_1 F_1}^{(1,0,0)}[\frac{5}{2},\frac{1}{2},-w^2])\nonumber\\
&&+|u|(-84+48\gamma_{E}+180u^2-96\gamma_{E}u^2-32u^4+24\gamma_{E}u^4+\nonumber\\
&&180w^2-96\gamma_{E}w^2-192u^2w^2+48\gamma_{E}u^2w^2-32w^4+\nonumber\\
&&24\gamma_{E}w^4+24i\mathbf{Erfi}[u]-9i\gamma_{E}\mathbf{Erfi}[u]-96iw^2\mathbf{Erfi}[u]+\nonumber\\
&&36i\gamma_{E}w^2\mathbf{Erfi}[u]+32iw^4\mathbf{Erfi}[u]-12i\gamma_{E}w^4\mathbf{Erfi}[u]+24{\log}[V_{T,s}]-\nonumber\\
&&24u^2{\log}[V_{T,s}]-24w^2{\log}[V_{T,s}]-18i\mathbf{Erfi}[u]{\log}[V_{T,s}]+72iw^2\mathbf{Erfi}[u]{\log}[V_{T,s}]-\nonumber\\
&&24iw^4\mathbf{Erfi}[u]{\log}[V_{T,s}]-6(-1+2u^2)(-1+2w^2)({_1 F_1}^{(1,0,0)}[0,\frac{1}{2},u^2]+{_1 F_1}^{(1,0,0)}[0,\frac{1}{2},w^2])-\nonumber\\
&&12e^{u^2}{_1 F_1}^{(1,0,0)}[\frac{3}{2},\frac{1}{2},-u^2]-12e^{w^2}{_1 F_1}^{(1,0,0)}[\frac{3}{2},\frac{1}{2},-w^2]+\nonumber\\
&&18e^{u^2}{_1 F_1}^{(1,0,0)}[\frac{5}{2},\frac{1}{2},-u^2]+18e^{w^2}{_1 F_1}^{(1,0,0)}[\frac{5}{2},\frac{1}{2},-w^2]+\nonumber\\
&&9ie^{w^2}\mathbf{Erfi}[u]{_1 F_1}^{(1,0,0)}[\frac{5}{2},\frac{1}{2},-w^2]+6{_1 F_1}^{(1,0,1)}[0,\frac{1}{2},u^2]-24u^2{_1 F_1}^{(1,0,1)}[0,\frac{1}{2},u^2]-\nonumber\\
&&12w^2{_1 F_1}^{(1,0,1)}[0,\frac{1}{2},u^2]+48u^2w^2{_1 F_1}^{(1,0,1)}[0,\frac{1}{2},u^2]+6{_1 F_1}[0,\frac{1}{2},w^2]-12u^2{_1 F_1}^{(1,0,1)}[0,\frac{1}{2},w^2]-\nonumber\\
&&24w^2{_1 F_1}^{(1,0,1)}[0,\frac{1}{2},w^2]+48u^2w^2{_1 F_1}^{(1,0,1)}[0,\frac{1}{2},w^2]+12u^2{_1 F_1}^{(1,0,2)}[0,\frac{1}{2},u^2]\nonumber\\
&&-24u^2w^2{_1 F_1}^{(1,0,2)}[0,\frac{1}{2},u^2]+\nonumber\\
&&12w^2{_1 F_1}^{(1,0,2)}[0,\frac{1}{2},w^2]-24u^2w^2{_1 F_1}^{(1,0,2)}[0,\frac{1}{2},w^2])\},
\end{eqnarray}
and
\begin{eqnarray}\label{eq:3.8}
&&\Upsilon(u,w)=\frac{-1}{8|u|}\{iu\mathbf{Erfi}[|u|]((-3+2w^2)(-2+\gamma_{E}+2{\log}[V_{T,s}])+e^{w^2}{_1 F_1}^{(1,0,1)}[\frac{3}{2},\frac{1}{2},-w^2])+\nonumber\\
&&|u|(22-12\gamma_{E}-8u^2+4\gamma_{E}u^2-8w^2+4\gamma_{E}w^2-\nonumber\\
&&6i\mathbf{Erfi}[u]+3i\gamma_{E}\mathbf{Erfi}[u]+4iw^2\mathbf{Erfi}[u]-2i\gamma_{E}w^2\mathbf{Erfi}[u]-4{\log}[V_{T,s}]+\nonumber\\
&&6i\mathbf{Erfi}[u]{\log}[V_{T,s}]-4iw^2\mathbf{Erfi}[u]{\log}[V_{T,s}]-2e^{u^2}{_1 F_1}^{(1,0,0)}[\frac{3}{2},\frac{1}{2},-u^2]-\nonumber\\
&&ie^{w^2}(-2i+\mathbf{Erfi}[u]){_1 F_1}^{(1,0,1)}[\frac{3}{2},\frac{1}{2},-w^2])\}.
\end{eqnarray}
Here, ${_1 F_1}[a;b;z]$ denoting Kummer's confluent hypergeometric function and the superscripts represent the derivative of the hypergeometric function with respect to its parameters, for example: ${_1 F_1}^{(1,0,0)}[a;b;z]$ represents the derivative with respect to the first parameter, i.e.~$a$, and $\mathbf{Erfi}[u]$ gives the imaginary error function $\mathbf{Erf}[iu]/i$.

\end{document}